\documentclass[twocolumn]{spie}  

\usepackage{amsmath,amsfonts,amssymb}
\usepackage{graphicx}
\usepackage{siunitx}
\usepackage[colorlinks=true, allcolors=blue]{hyperref}
\usepackage{booktabs}
\usepackage{multirow}
\usepackage{color,soul}
\usepackage{array}
\newcolumntype{P}[1]{>{\centering\arraybackslash}p{#1}}
\newcolumntype{M}[1]{>{\centering\arraybackslash}m{#1}}
\title{Video compression with low complexity CNN-based spatial resolution adaptation}

\author[a]{Di Ma, Fan Zhang and David R. Bull}

\affil[a]{Bristol Vision Institute, University of Bristol, Bristol, United Kingdom}

\authorinfo{Correspondence emails: di.ma@bristol.ac.uk, fan.zhang@bristol.ac.uk, and dave.bull@bristol.ac.uk}

\begin{document} 
\maketitle

\begin{abstract}

It has recently been demonstrated that spatial resolution adaptation can be integrated within video compression to improve overall coding performance by spatially down-sampling before encoding and super-resolving at the decoder. Significant improvements have been reported when convolutional neural networks (CNNs) were used to perform the resolution up-sampling. However, this approach suffers from high complexity at the decoder due to the employment of CNN-based super-resolution. In this paper, a novel framework is proposed which supports the flexible allocation of complexity between the encoder and decoder. This approach employs a CNN model for video down-sampling at the encoder and uses a Lanczos3 filter to reconstruct full resolution at the decoder. The proposed method was integrated into the HEVC HM 16.20 software and evaluated on JVET UHD test sequences using the All Intra configuration. The experimental results demonstrate the potential of the proposed approach, with significant bitrate savings (more than 10\%) over the original HEVC HM, coupled with reduced computational complexity at both encoder (29\%) and decoder (10\%).

\end{abstract}

\keywords{Spatial resolution adaptation, convolutional neural networks, CNN-based spatial down-sampling, video compression, HEVC}

\section{INTRODUCTION}
\label{sec:intro}  

Deep learning techniques, especially using convolutional neural networks (CNNs), have provided powerful solutions to many image and video processing problems, such as classification, detection, enhancement and super-resolution \cite{hemanth2017deep}. More recently, deep learning has also been integrated with image and video compression to build end-to-end coding frameworks \cite{balle2016end,lu2019dvc,djelouah2019neural} or to enhance coding modules in standard coding algorithms \cite{cui2018convolutional,zhang2019advanced,zhang2020pp,ma2020mfrnet}. These approaches demonstrate significant potential for improvements in coding efficiency. 

One of the CNN-based coding tools which can offer significant coding gains is resolution adaptation. This spatially down-samples image (or video frame) resolution using a simple filter (e.g. Bicubic or Lanczos3) before encoding and employs a CNN-based up-sampling approach to restore the original resolution at the decoder. An issue with this type of method is the high computational complexity introduced in the decoding process due to the CNN up-sampling operation. Recently a CNN-based down-sampling method \cite{li2018learning} for spatial resolution adaptation has been proposed in the context of image compression. However this method still requires a CNN-based up-sampling approach applied at the decoder, without achieving complexity reduction at the encoder.

In this paper, we propose a low complexity CNN-based spatial resolution adaptation framework for video compression. This employs a CNN model for down-sampling at the encoder and a simple Lanczos3 up-sampling filter to generate full resolution video frames during decoding. This approach offers a trade-off solution between computational complexity and coding performance, and enables flexible complexity allocation between the encoder and decoder. The proposed method has been integrated into the HEVC HM 16.20 test model and evaluated on the JVET Ultra High Definition (UHD) test sequences using the All Intra configuration. Significant coding gains (more than 10\%)  have been achieved by this approach when compared to the unmodified HEVC HM 16.20,  coupled with reduced complexity for both encoding and decoding processes. It also provides evident coding efficiency improvement compared to using simple filters for both down- and up-sampling operations.

The remainder of the paper is organised as follows. Section \ref{sec:algorithm} describes the proposed spatial resolution adaptation framework, while Section \ref{sec:results} presents the evaluation results for both compression performance and complexity analysis. Finally, conclusions and future work are provided in Section \ref{sec:conclusion}.

\section{Proposed Algorithm}
\label{sec:algorithm}

\begin{figure*}[ht]
\centering
\includegraphics[width=12cm]{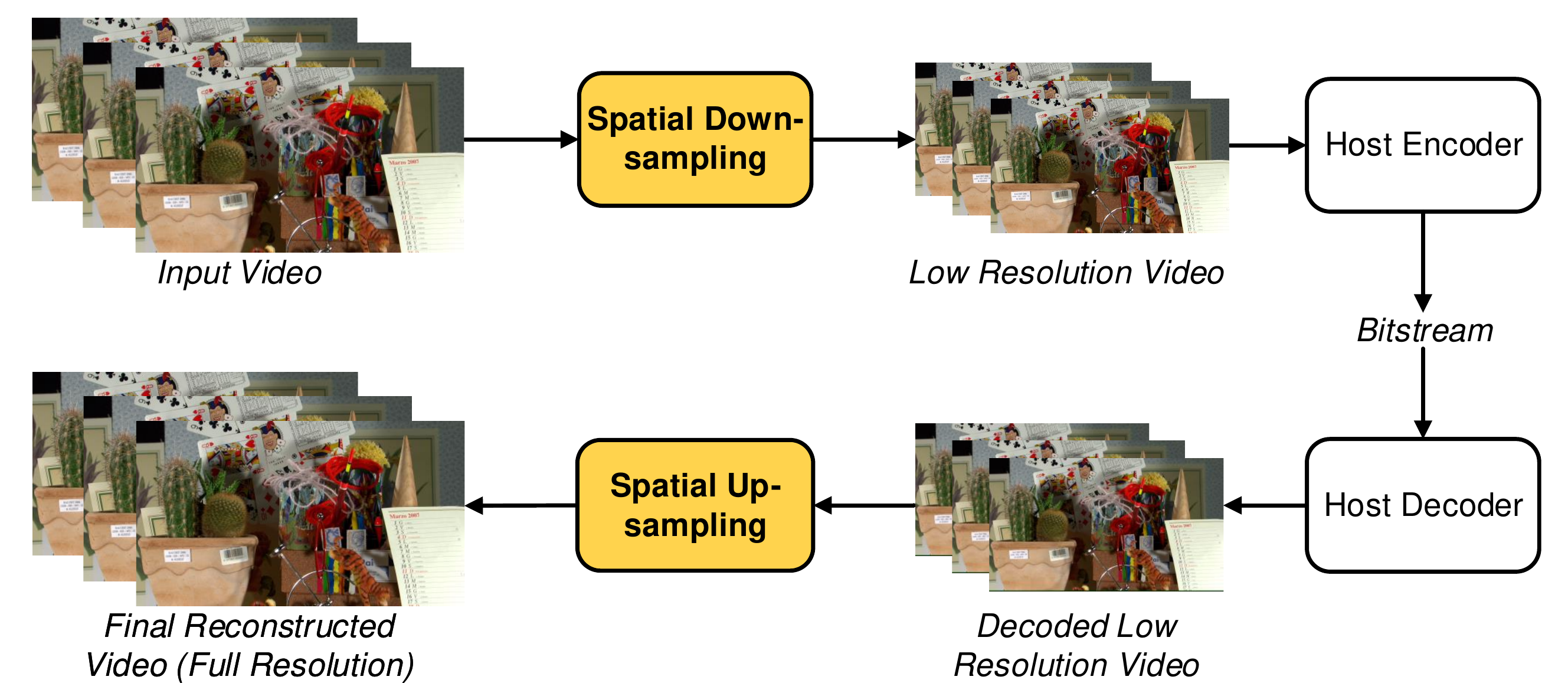}
\caption{Diagram of the generic spatial resolution adaptation workflow. \label{fig:framework}}
\end{figure*}

A generic spatial resolution adaptation (SRA) framework for video compression is illustrated in Figure \ref{fig:framework}. According to the various possible approaches employed in spatial down-sampling and up-sampling, there exist four different scenarios:

\begin{itemize}
    \item \textbf{Scenario 1}: Simple filters (e.g. Lanczos3) are used for both down-sampling and up-sampling.
    \item \textbf{Scenario 2}: A CNN model is utilised for down-sampling, and up-sampling is achieved through simple filtering.
        \item \textbf{Scenario 3}: A simple filter is employed for down-sampling, while a CNN-based super-resolution approach is used for up-sampling.
    \item \textbf{Scenario 4}: Both down-sampling and up-sampling processes are CNN-based.
\end{itemize}

It is noted that Scenarios 1, 3 and 4 have been previously investigated \cite{afonso2017low,afonso2019video,li2018learning} in the literature. However we are not aware of successful examples based on Scenario 2. This is primarily due to the difficulty of achieving end-to-end training for a down-sampling CNN, especially when video compression is included in the workflow \cite{li2018learning}.

In this work we address Scenario 2, proposing a new CNN architecture for spatial resolution down-sampling in the context of video compression. The employed training strategy is described in this section alongside the database and loss function employed. 

\subsection{The employed CNN architecture}

\begin{figure*}[ht]
\centering
\includegraphics[width=17.2cm]{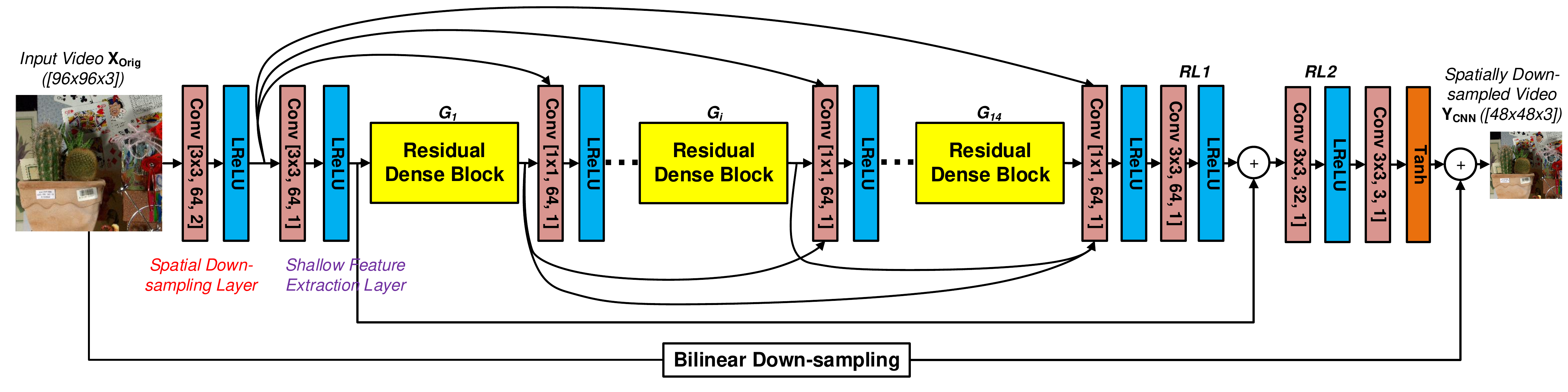}
\caption{Network architecture of the proposed DSNet.\label{fig:generator}}
\end{figure*}

Existing CNN-based SRA approaches often employ simple filters (e.g. Bicubic or Lanczos3) to achieve spatial resolution down-sampling. These filters have constant parameters and may result in the loss of important spatial information during the down-sampling process, which may be beneficial to the up-sampling operation. In this work, a Deep Down-sampling CNN, DSNet, is developed for down-sampling aimed at improvements in overall coding performance while reducing overall complexity. The architecture of this network is shown in Figure \ref{fig:generator}.

\begin{figure}[h]
\centering
\includegraphics[width=0.925\linewidth]{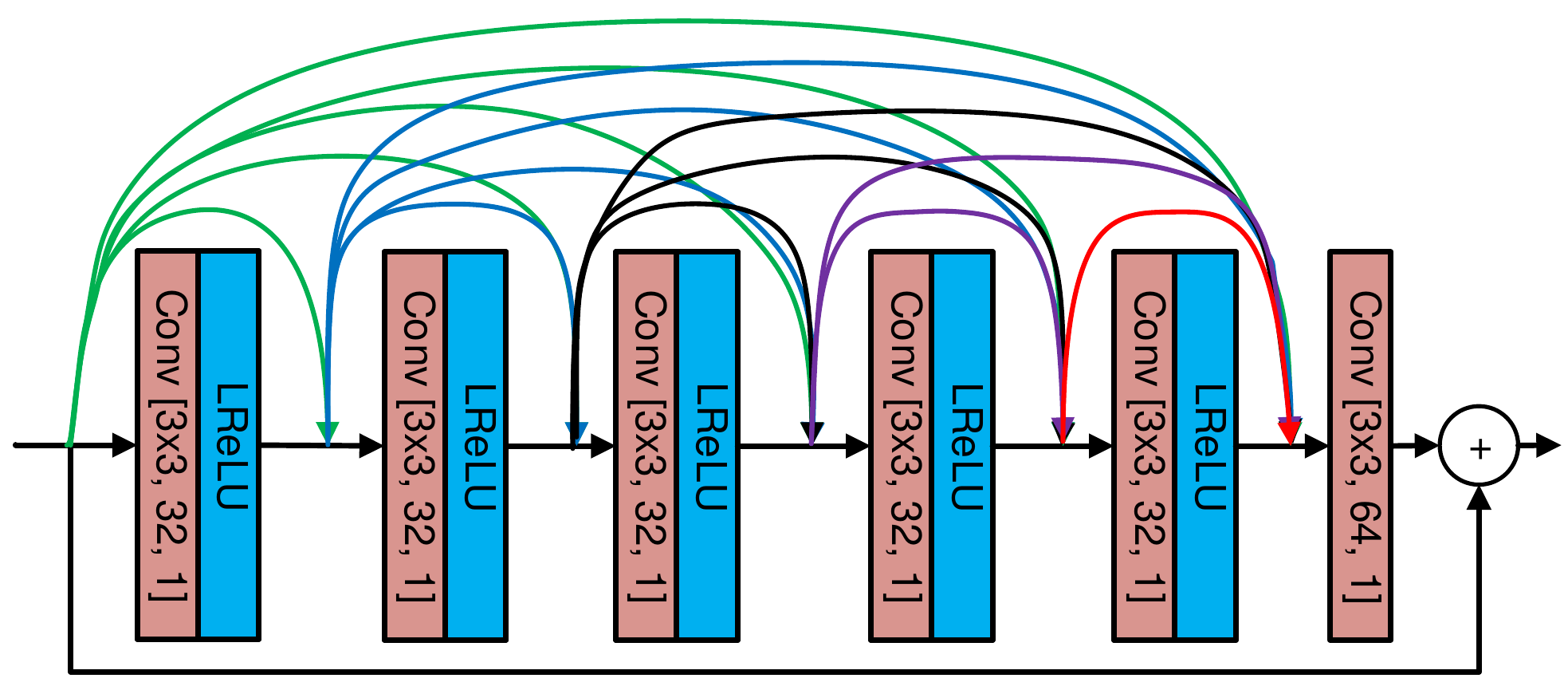}
\caption{Residual Dense Block (RDB) used in DSNet.\label{fig:RDB}}
\end{figure} 

This network is based on a modified version of our own CNN architecture developed for bit depth up-sampling \cite{ma2020gan}. It takes a 96$\times$96 YCbCr (4:4:4) image block as input, and outputs a low resolution block (48$\times$48) in the same format. The input signal is first processed by a shallow spatial down-sampling layer and a feature extraction layer, each of which consists of a convolutional layer and a Leaky ReLU (LReLU) activation function. After the shallow feature extraction layer, 14 residual dense blocks (RDBs) \cite{zhang2018residual} were employed to further extract dense features. Multiple cascading connections (shown as black curves in Figure \ref{fig:generator}) are designed to connect these 14 RDBs and feed the outputs of the spatial down-sampling layer and each RDB ($G_{i}$, $i=1,2,...,13$) into the subsequent RDBs or the first reconstruction layer (RL1) through a 1$\times$1 convolutional layer with a LReLU activation function. A skip connection is further utilised to connect the outputs of the shallow feature extraction layer and RL1. Another reconstruction layer (RL2) is employed which is followed by the final convolution layer to produce the residual signal. Finally, the input image block is spatially down-sampled by a factor of 2 using a Bilinear filter and combined with the residual signal using a long skip connection to output the final down-sampled image block. The number of feature maps, kernel sizes and stride values for all convolutional layers are presented in Figure \ref{fig:generator}.

Figure \ref{fig:RDB} illustrates the structure of each RDB employed in the proposed network. Due to the dense connection, the convolutional layer in each RDB fully reuses features from its preceding layers \cite{huang2017densely}, which effectively improves information flow between these layers. An additional skip connection is also designed to connect the input and output of each RDB in order to stabilise training and evaluation processes \cite{he2016deep}.

\subsection{Training Database}
\label{sub:setup}

Extensive and diverse training data is essential for CNN-based video compression, in order to optimise model generalisation and prevent overfitting problems \cite{tian2018tdan}. To effectively train the proposed network, our large and representative database BVI-DVC \cite{ma2020bvi} is employed to generate training material. This database contains 800 10 bit YCbCr 4:2:0 video sequences at four different spatial resolutions from 270p to 2160p, including various content and texture types. The video frames of these sequences were randomly selected, segmented into 96$\times$96 image blocks, and converted into YCbCr 4:4:4 format. During this process, block rotation is applied to achieve data argumentation. This results in approximately 200,000 image blocks in total. They are employed as both input and the training target of the CNN.

\subsection{Loss Function}
\label{sub:loss}

Loss functions are a key component in CNN training. Since the proposed network is used for resolution down-sampling before encoding, the CNN output (at low resolution) should preserve sufficient spatial information to enble high fidelity full resolution reconstruction (up-sampling) at the decoder. On the other hand, too much high frequency information may lead to higher bitrates during compression. This results in an optimisation problem which is similar to traditional rate distortion optimisation \cite{bull2014communicating,zhang2019RDO} in image and video coding. 

To solve this problem, the ideal solution would be to conduct an end-to-end optimisation which includes an image or video codec in the training loop -- to encode the low resolution CNN output and generate a real bitstream whose corresponding bitrate ($R$) could be measured. The decoded low-resolution image block could then be up-sampled using a simple filter (e.g. Lanczos3) to get the final reconstructed full resolution content for comparison with  its original (uncompressed full resolution) counterpart to calculate overall distortion ($D$). The loss function ($\mathcal L$) used during CNN-training can be designed as equation (\ref{eq:RDO}) employing the Lagrange multiplier method.
\begin{equation}
   \mathcal L = D(\mathbf{p}) + \lambda_\mathrm{CNN} \cdot R(\mathbf{p}).
    \label{eq:RDO}
\end{equation}
Here  $\mathbf{p}$ represents the CNN parameters which need to be optimised during training. $\lambda_\mathrm{CNN}$ is the Lagrange multiplier which is used to trade off the relationship between $D$ and $R$. 

In practice, we note that conventional image or video codecs, such as HEVC HM, cannot to be integrated into the training loop due to incompatibility with existing machine learning libraries (e.g. TensorFlow and Pytorch) \cite{li2018learning}. In order to optimise the proposed network and achieve superior overall rate quality performance (based on the framework in Scenario 2), a loss function is proposed to emulate the rate distortion optimisation process, as shown in equation (\ref{eq:2}).    
\begin{equation}
\begin{aligned}
     \mathcal L_{\rm DSNet}= \rm MSE(\rm X_{Orig},\rm Y_{CNN}\__{BicUp}) + \lambda_\mathrm{DSNet} \cdot \\ 
     (\rm MSE(\rm Y_{L3},\rm Y_{CNN}) + \omega \cdot \mathcal L_{\rm MS-SSIM}(\rm Y_{L3},\rm Y_{CNN})).
     \end{aligned}
\label{eq:2}
\end{equation}
The first term $\rm MSE(\rm X_{Orig},\rm Y_{CNN}\__{BicUp})$ calculates the mean squared error (MSE) between the original full resolution input block ($\rm X_{Orig}$) and the Bicubic up-sampled CNN output ($\rm Y_{CNN}\__{BicUp}$). This accounts for the distortion generated during the resolution adaptation process.  $\rm MSE(\rm Y_{L3},\rm Y_{CNN})$ represents the MSE between the CNN output low resolution image block ($\rm Y_{CNN}$) and the Lanczos3 filter down-sampled (from the original) low resolution image block ($\rm Y_{L3}$), and their MS-SSIM \cite{wang2003multiscale} loss is also obtained in the term $\mathcal L_{\rm MS-SSIM}(\rm Y_{L3},\rm Y_{CNN})$. The weighted linear combination between $\rm MSE(\rm Y_{L3},\rm Y_{CNN})$ and $\mathcal L_{\rm MS-SSIM}(\rm Y_{L3},\rm Y_{CNN})$ is employed to estimate the bitrate level when the CNN low resolution output is compressed. $\omega$ and $\lambda_\mathrm{DSNet}$ are two constant parameters representing the weights used in the combination model and the Lagrange multiplier respectively.

\subsection{Training and Evaluation Configurations}

The proposed DSNet was implemented and trained based on the TensorFlow (version 1.8.0) framework with the following training parameters: Adam optimisation \cite{kingma2014adam} with hyper-parameters of $\beta_1$=0.9 and $\beta_2$=0.999; batch size of 4$\times$4; 200 training epochs; learning rate  of 0.0001; weight decay of 0.1 for every 100 epochs. The used parameter values for the Lagrange multiplier $\lambda_\mathrm{DSNet}$ and the weight $\omega$ were 30 and 1/6 respectively which have been determined based on a  ten-fold cross-validation using the BVI-DVC database \cite{ma2020bvi}.

During network evaluation, each full resolution frame of the test sequence is segmented into 96$\times$96 overlapping blocks with an overlap size of 8 pixels, and converted to YCbCr 4:4:4 format as network input. The network output image blocks (with size of 48$\times$48) are then converted to the original format (YCbCr 4:2:0) and aggregated in the same way (overlap size equals 4 pixels) to generate the spatially down-sampled video frame.
 
\section{Results and Discussions}
\label{sec:results}

The proposed spatial down-sampling CNN architecture has been integrated into the spatial resolution adaptation (SRA) framework (Scenario 2) and fully evaluated with HEVC HM 16.20 as the host codec. The evaluation followed the All Intra (Main 10 Profile) configuration used in the JVET Common Test Conditions \cite{bossen2018jvet} using six JVET UHD sequences as test material. None of these test sequences were used for training the proposed CNN model. Four initial base quantisation parameter (QP) values are employed: 27, 32, 37 and 42. 

In order to achieve similar bitrate ranges and hence a meaningful comparison between the proposed approach and the original HEVC, a fixed QP offset of -6 is applied on the base QP value during encoding when SRA is enabled \cite{afonso2019video}. We have also noted that the coding improvement achieved by spatial resolution adaptation is highly content dependent. For some sequences at certain QP values, SRA may not offer coding gains over the original host codec. Therefore we have employed a quantisation resolution optimisation (QRO) module \cite{zhang2019vistra2}, which employs a machine learning based approach to make decisions on resolution adaptation based on a spatial resolution dependent quality metric, SRQM \cite{mackin2018srqm}, temporal information (TI) \cite{winkler2012analysis} and initial base QP values. For cases when spatial resolution adaptation is not activated, the test sequences will be compressed using the original HEVC HM with the initial base QP. 

To benchmark the coding performance of SRA Scenario 2, we have generated results for SRA Scenarios 1, 3 and 4 alongside original HEVC compression (HM 16.20). For simple filter-based down- and/or up-sampling in Scenario 1 and 3, we have used Lanczos3 filters. In Scenario 3 and 4, a previously developed super-resolution CNN, MSRResNet \cite{ma2019perceptually,zhang2019enhanced} is employed for resolution up-sampling at the decoder.

\subsection{Compression Performance}

\begin{table*}[!t]
\centering
\textbf{Table 1: Compression performance comparison between various SRA scenarios and the original HEVC HM 16.20 (AI configuration) (``$\downarrow$'' and ``$\uparrow$'' represent spatial down-sampling and up-sampling respectively).}
\begin{tabular}{l |M{2.8cm}|M{2.8cm}|M{2.8cm}|M{2.8cm}}
\toprule
\multirow{3}{*}{Sequence} & Scenario 1 & Scenario 2 & Scenario 3 & Scenario 4 \\
&{L3 $\downarrow$ \& L3 $\uparrow$} & 
{CNN $\downarrow$ \& L3 $\uparrow$}&{L3 $\downarrow$ \& CNN $\uparrow$}&
{CNN $\downarrow$ \& CNN $\uparrow$}\\
\cmidrule{2-5}
\centering
&   BD-rate (PSNR) &    BD-rate (PSNR)&BD-rate (PSNR)&    BD-rate(PSNR) \\
\midrule \midrule
Campfire&-7.4\% & -8.8\% &-17.8\% &-18.6\%\\
FoodMarket4&-7.8\% & -8.4\% & -12.3\%&-13.6\%\\
Tango2&-9.7\% & -10.6\% & -12.8\%& -14.8\% \\
CatRobot1&-4.4\% & -5.2\% & -13.0\%&-14.7\%\\
DaylightRoad2&-2.6\% & -3.8\% &-7.9\% &-10.2\%\\
ParkRunning3&-23.6\% & -24.5\% &-26.9\% &-28.2\%\\
\midrule \textbf{Average} & -9.2\% & -10.2\% &  -15.1\%& -16.7\%
\\\bottomrule
\end{tabular}
	\end{table*}

\begin{table*}[htbp]
\centering
\textbf{Table 2: Relative complexity for four SRA Scenarios.}
\begin{tabular}{l| M{1.2cm}| M{1.2cm}|M{1.2cm}| M{1.2cm}| M{1.2cm}|M{1.2cm}| M{1.2cm}|M{1.2cm}}

\toprule
\multirow{3}{*}{Sequence} & \multicolumn{2}{c|}{Scenario 1} & \multicolumn{2}{c|}{Scenario 2} & \multicolumn{2}{c|}{Scenario 3} & \multicolumn{2}{c}{Scenario 4} \\
&\multicolumn{2}{c|}{L3 $\downarrow$ \& L3 $\uparrow$} & \multicolumn{2}{c|}{CNN $\downarrow$ \& L3 $\uparrow$}&\multicolumn{2}{c|}{L3 $\downarrow$ \& CNN $\uparrow$}& \multicolumn{2}{c}{CNN $\downarrow$ \& CNN $\uparrow$}\\
\cmidrule{2-9}
\centering
&   Enc &    Dec &   Enc &    Dec& Enc & Dec &Enc&Dec\\
 \midrule \midrule
 Campfire& 0.47$\times$ & 0.96$\times$&0.83$\times$&0.95$\times$&0.47$\times$&29.5$\times$&0.83$\times$&29.3$\times$\\
 FoodMarket4 &0.23$\times$&0.71$\times$&0.54$\times$&0.68$\times$&0.23$\times$&25.4$\times$&0.54$\times$&25.2$\times$\\
 Tango2 &0.32$\times$&0.98$\times$&0.52$\times$&0.95$\times$&0.32$\times$&30.5$\times$&0.52$\times$&30.2$\times$\\
 CatRobot1 &0.53$\times$&1.20$\times$&0.85$\times$&1.10$\times$&0.53$\times$&34.3$\times$&0.85$\times$&34.1$\times$\\
DaylightRoad2 &0.34$\times$&0.98$\times$&0.57$\times$&0.94$\times$&0.34$\times$&29.2$\times$&0.57$\times$& 29.0$\times$\\
ParkRunning3 &0.73$\times$&0.89$\times$&0.94$\times$&0.86$\times$&0.73$\times$&26.9$\times$&0.94$\times$& 26.5$\times$\\
\midrule 
Average &0.44$\times$&0.95$\times$&0.71$\times$&0.90$\times$&0.44$\times$&29.3$\times$&0.71$\times$&29.1$\times$
\\\bottomrule
\end{tabular}
\label{tab:complexity}
\end{table*}
	
Table 1 summaries the compression performance for the four different SRA approaches (Scenario 1-4) when they are compared to that of the original HEVC HM 16.20 using Bj{\o}ntegaard Delta \cite{BD} measurement (BD-rate) based on the assessment of PSNR (Peak to Noise-Signal-Ratio, luminance channel only). It can be observed that when a CNN is employed for resolution down-sampling (Scenario 2 and 4), additional coding gains have been achieved over those scenarios where Lanczos3 down-sampling is applied (Scenario 1 and Scenario 3 respectively). This improvement is consistent among all six test sequences. We have also noticed that when CNN-based super-resolution is utilised (Scenario 3 and 4), the bitrate savings are more significant (up to 16.7\%) against Scenario 1 and 2 with Lanczos3 up-sampling. 


\subsection{Complexity Analysis}

The encoder and decoder complexities for all five evaluated methods (HM 16.20 and SRA Scenario 1-4) were also calculated. The encoding was executed on a shared cluster, BlueCrystal Phase 3 \cite{bc3}, based at the University of Bristol, which has 223 base blades. Each blade contains 16 2.6GHz SandyBridge cores and 64GB RAM. The decoding was conducted on a PC with an Intel(R) Core(TM) i7-4770K CPU @3.5GHz, 24GB RAM and NVIDIA P6000 GPU device. The encoding and decoding execution times for SRA Scenario 1-4 are all benchmarked against those for the original HEVC HM.

Table 2 reports the average (for the four evaluated QPs) relative encoding and decoding complexities of the four SRA scenarios for six test sequences. When a Lanczos3 filter is used for down-sampling (Scenario 1 and 3), the encoding complexity is only 44\% of that for the original HM. This is due to the simplicity of the down-sampling filter and the encoding of low resolution content. The proposed CNN-based down-sampling in Scenario 2 and 4 can also reduce the overall encoding time by approximately 30\% and offers better overall rate quality performance compared to Lanczos3 down-sampling (as shown in Table 1). It is also noted that the decoding complexity has also been slightly reduced in Scenario 2 and 4 (with CNN based down-sampling) compared to Scenario 1 and 3 (based on Lanczos3 down-sampling) respectively. This may be because CNN-based down-sampling generated content is relatively easy to compress.

As mentioned in Section \ref{sub:loss}, the employed training strategy is a sub-optimal solution. The results presented in this section demonstrate its potential, while acknowledging that and overall coding performance can be further enhanced if a more realistic end-to-end optimisation is applied. Our approach is particularly relevant to application scenarios where there are limited resource available at the decoder or where CNN-based up-sampling cannot be supported. In such cases, we have shown that consistent coding gains can still be achieved by re-distributing the computational complexity from the decoder to the encoder (as in Scenario 2) by applying CNN-based down-sampling instead of CNN-based super-resolution.




\section{Conclusions}
\label{sec:conclusion}

In this paper, a low complexity CNN-based spatial resolution adaptation framework is proposed for video compression. This method employs a CNN-based down-sampling approach before encoding and applies up-sampling at the decoder using a simple filter. The proposed approach has been integrated with HEVC HM 16.20 and evaluated on JVET-CTC UHD test sequences (under the All Intra configuration). Improved coding performance has been achieved compared to the original HEVC HM and against Lanczos3 filter based re-sampling, coupled with reduced computational complexity at both encoder and decoder. Our future work will continue to enhance the CNN training methodology, extend the approach to inter-coding configurations and optimise our CNN models for different quantisation level ranges.

\acknowledgments 
 
The authors acknowledge funding from EPSRC (EP/L016656/1 and EP/M000885/1) and the NVIDIA GPU Seeding Grants.

\bibliographystyle{spiebib} 
\bibliography{refs} 
\end{document}